\documentclass[12pt, draftclsnofoot, onecolumn]{IEEEtran}

\usepackage{amssymb}
\usepackage{amsmath}
\usepackage{cite}
\usepackage{url}
\usepackage{empheq}
\usepackage{xcolor}
\usepackage{graphicx}
\usepackage{subfigure}
\usepackage{enumitem}
\usepackage{fancyhdr}
\usepackage{mdwmath}	
\usepackage{mdwtab}
\usepackage{caption}
\usepackage{amsthm}
\usepackage{algorithm}
\usepackage{algorithmic}

\newtheorem{lemma}{Lemma}

\newtheorem{theorem}{Theorem}
\newtheorem{corollary}{Corollary}
\newtheorem{assumption}{Assumption}
\newtheorem{definition}{Definition}

\newcommand{\eqr}[1]{(\ref{#1})}
\newcommand{\fref}[1]{Fig.~\ref{#1}}

\begin{document}
\title{Pinching-Antenna Systems with LoS Blockages}
\author{Kaidi~Wang,~\IEEEmembership{Member,~IEEE,}
Chongjun~Ouyang,~\IEEEmembership{Member,~IEEE,}
Yuanwei~Liu,~\IEEEmembership{Fellow,~IEEE,}
and Zhiguo~Ding,~\IEEEmembership{Fellow,~IEEE}
\thanks{K. Wang and Z. Ding are with the Department of Electrical and Electronic Engineering, the University of Manchester, M1 9BB Manchester, U.K. (email: kaidi.wang@ieee.org; zhiguo.ding@ieee.org).}
\thanks{C. Ouyang is with the School of Electronic Engineering and Computer Science, Queen Mary University of London, London, E1 4NS, U.K. (email: c.ouyang@qmul.ac.uk).}
\thanks{Y. Liu is with the Department of Electrical and Electronic Engineering, The University of Hong Kong, Hong Kong (email: yuanwei@hku.hk).}
\thanks{Z. Ding is also with Khalifa University, Abu Dhabi, UAE.}}
\maketitle
%%%%%%%%%%%%%%%%%%%%%%%%%%%%%%%%%%%%%%%%%%%%%%%%%
%%%%%%%%%%%%%%%%%%%%%%%%%%%%%%%%%%%%%%%%%%%%%%%%%
%\setlength{\abovedisplayskip}{2pt}
%\setlength{\belowdisplayskip}{2pt}
%\setlength{\textfloatsep}{0pt}
\begin{abstract}
The aim of this letter is to explore the capability of pinching-antenna systems to construct line-of-sight (LoS) links in the presence of LoS blockages. Specifically, pinching antennas are pre-installed at preconfigured positions along waveguides and can be selectively activated to create LoS links for enhancing desired signals and non-line-of-sight (NLoS) links for eliminating inter-user interference. On this basis, a sum-rate maximization problem is formulated by jointly optimizing waveguide assignment and antenna activation. To solve this problem, a matching based algorithm is proposed using two distinct preference designs. Simulation results demonstrate that the considered pinching-antenna system and proposed solutions can dynamically establish LoS links and effectively exploit LoS blockages to mitigate interference, thereby significantly improving system throughput.
\end{abstract}
\vspace{-4mm}
%%%%%%%%%%%%%%%%%%%%%%%%%%%%%%%%%%%%%%%%%%%%%%%%% 
%%%%%%%%%%%%%%%%%%%%%%%%%%%%%%%%%%%%%%%%%%%%%%%%%
\section{Introduction}
The advancement of sixth-generation (6G) wireless systems has increased the demand for advanced antenna technologies that can deliver high data rates, reliability, and adaptability. In response, flexible antennas, including fluid antennas and movable antennas, have been proposed to provide reconfigurability beyond the limitations of conventional fixed-location designs \cite{wong2020fluid, zhu2023modeling}. Although these technologies enable positional adjustments through distinct mechanisms, they face challenges in establishing reliable line-of-sight (LoS) transmission paths, particularly in environments with physical obstructions where wavelength-scale movements prove insufficient. To address these challenges, pinching antennas have been introduced recently \cite{ding2024pin, ding2025blockage}. By placing dielectric particles along a waveguide, pinching antennas can dynamically construct LoS paths and enhance transceiver channels, thereby improving system performance in obstructed environments \cite{suzuki2022pinching, liu2025pinching}.

To improve the performance of pinching-antenna systems, extensive optimization studies have been conducted \cite{xu2025pin2, xu2025joint, kaidi2025pin2}. In \cite{xu2025pin2}, a two-user scenario was studied where the data rate of one user is maximized while ensuring the quality-of-service (QoS) of the other through joint optimization of antenna placement and power allocation. The authors of \cite{xu2025joint} investigated a multi-waveguide configuration, focusing on the joint design of antenna placement and beamforming by developing both optimization based and learning based methods. More recently, \cite{kaidi2025pin2} proposed a practical implementation of multi-waveguide systems with pre-installed pinching antennas, solving a comprehensive optimization problem that includes waveguide assignment, antenna activation, and power allocation. 

Although pinching-antenna systems have been extensively studied, their potential for constructing LoS links has not been exploited, which motivates this work. A recent research has proven that in the presence of LoS blockages, the superiority of pinching-antenna systems becomes more significant \cite{ding2025blockage}. Building on this insight, this work incorporates multiple LoS blockages and proposes a method to establish LoS links by selectively activating pre-installed pinching antennas across different waveguides. With the aim of maximizing the sum rate, waveguide assignment and antenna activation are jointly formulated and solved using matching theory, utilizing two different preference designs to balance performance and complexity. Simulation results demonstrate significant advantages of the pinching-antenna system over conventional fixed-location antenna systems in establishing LoS links, and show that the LoS blockages can be effectively exploited to suppress interference, thereby improving system throughput.
%%%%%%%%%%%%%%%%%%%%%%%%%%%%%%%%%%%%%%%%%%%%%%%%%
%%%%%%%%%%%%%%%%%%%%%%%%%%%%%%%%%%%%%%%%%%%%%%%%%
\section{System Model and Problem Formulation}
\begin{figure}[!t]
\centering{\includegraphics[width=105mm]{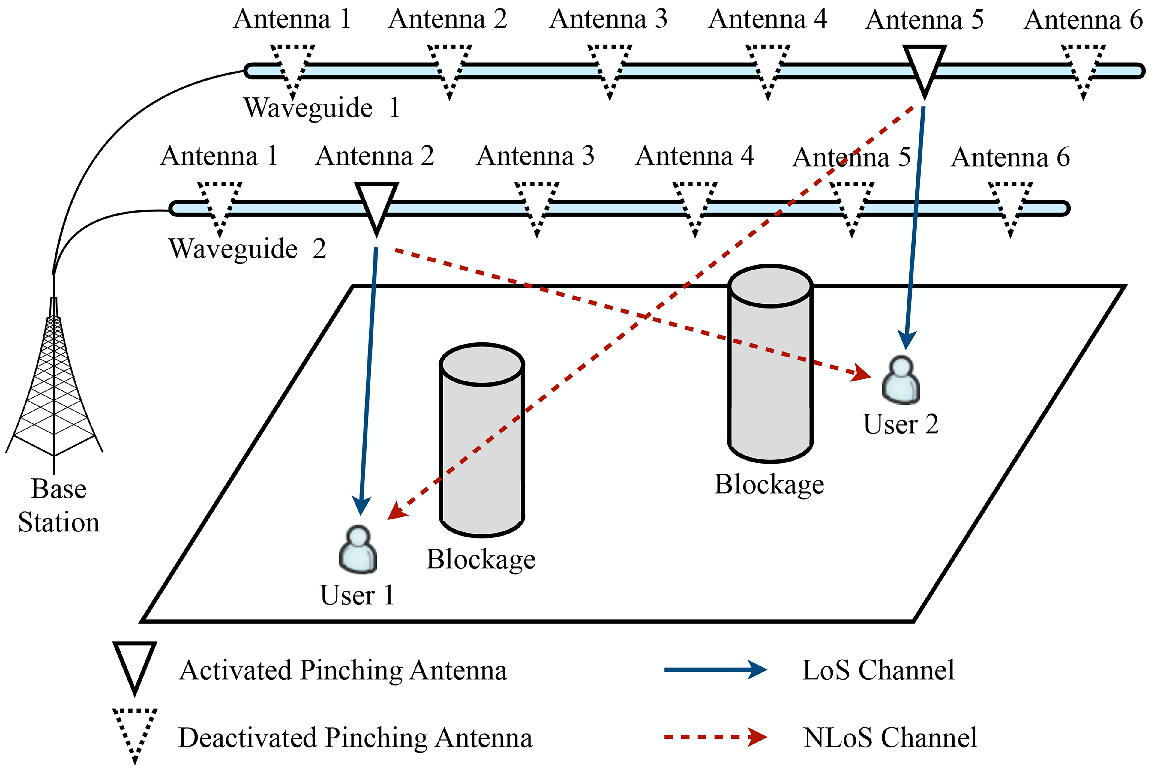}}
\caption{An illustration of the considered multi-waveguide pinching-antenna system with $N=2$ users, $M=6$ pinching antennas, and $K=2$ waveguides.}
\label{system}
\vspace{-4mm}
\end{figure}

A downlink multi-waveguide pinching-antenna system is considered in the presence of LoS blockages, where a base station simultaneously serves $N$ single-antenna users via $K$ parallel waveguides, with $N=K$. As shown in Fig.~\ref{system}, the coverage area is modeled as a rectangular region with dimensions $D_x \times D_y$, within which the LoS blockages are located at fixed positions. The waveguides are deployed at a height of $d$ and extend along the $x$-axis over a length of $D_x$. Each waveguide is equipped with $M$ pinching antennas installed at predetermined discrete positions. In the considered system, each user is assigned to a unique waveguide, and each waveguide transmits only the desired signal intended for the corresponding user. Furthermore, on each waveguide, a single pinching antenna can be activated for transmission.

The sets of all users, waveguides, and pinching antennas on waveguide $k$ are denoted by $\mathcal{N}\triangleq\{1,2,\cdots,N\}$, $\mathcal{K}\triangleq\{1,2,\cdots,K\}$, and $\mathcal{M}_k\triangleq\{1,2,\cdots,M\}$, respectively. The location of user $n$ is denoted by $\boldsymbol{\psi}_n=\left(x_n,y_n,0\right)$, and the location of pinching antenna $m$ on waveguide $k$ is represented by $\boldsymbol{\psi}_{k,m}^\mathrm{pin}=(x_{k,m}^\mathrm{pin},y_k^\mathrm{pin},d)$.
%%%%%%%%%%%%%%%%%%%%%%%%%%%%%%%%%%%%%%%%%%%%%%%%%
\subsection{Mixed LoS and NLoS Channel}
In the presence of a LoS link, based on the spherical wave channel model, the LoS path component between pinching antenna $m$ on waveguide $k$ and user $n$ is given by
\begin{equation}
h_{k,m,n}^\mathrm{LoS} = \frac{\eta e^{-j\frac{2\pi}{\lambda}\left\|\boldsymbol{\psi}_n-\boldsymbol{\psi}_{k,m}^\mathrm{pin}\right\|}}{\left\|\boldsymbol{\psi}_n-\boldsymbol{\psi}_{k,m}^\mathrm{pin}\right\|},
\end{equation}
where $\eta=\frac{c}{4\pi f_c}$ is the free-space path loss coefficient, $c$ is the speed of light, $f_c$ is the carrier frequency, $\lambda$ is the carrier wavelength, and $\|\boldsymbol{\psi}_n-\boldsymbol{\psi}_{k,m}^\mathrm{pin}\|$ is the Euclidean distance between user $n$ and pinching antenna $m$ on waveguide $k$. The non-line-of-sight (NLoS) path component is modeled as follows \cite{liu2023nf}:
\begin{equation}
h_{k,m,n}^\mathrm{NLoS} = \sum_{\ell=1}^{L}\delta_\ell\frac{\eta e^{-j\frac{2\pi}{\lambda} \left(\left\|\boldsymbol{\psi}_n-\boldsymbol{\psi}_\ell^\mathrm{scat}\right\|+\left\| \boldsymbol{\psi}_{k,m}^\mathrm{pin}-\boldsymbol{\psi}_\ell^\mathrm{scat}\right\|\right)}}{\left\|\boldsymbol{\psi}_n-\boldsymbol{\psi}_\ell^\mathrm{scat}\right\|\left\| \boldsymbol{\psi}_{k,m}^\mathrm{pin}-\boldsymbol{\psi}_\ell^\mathrm{scat}\right\|},
\end{equation}
where $L$ is the number of scatterers, $\delta_\ell$ is the complex gain of scattering path $\ell$, and $\boldsymbol{\psi}_\ell^\mathrm{scat}=(x_\ell^\mathrm{scat},y_\ell^\mathrm{scat},0)$ is the location of scatterer $\ell$.

By introducing a binary LoS indicator $\phi_{k,m}^n\in\{0,1\}$, the channel between pinching antenna $m$ on waveguide $k$ and user $n$ can be expressed as follows:
\begin{equation}
h_{k,m,n} = \phi_{k,m}^nh_{k,m,n}^\mathrm{LoS}+h_{k,m,n}^\mathrm{NLoS},
\end{equation}
where $\phi_{k,m}^n=1$ indicates the existence of a LoS path between user $n$ and pinching antenna $m$ on waveguide $k$, and $\phi_{k,m}^n=0$ otherwise. By incorporating antenna activation, one pinching antenna is activated on each waveguide, and the effective channel between waveguide $k$ and user $n$ is given by
\begin{equation}
h_{k,n} = \sum_{m=1}^M \beta_{k,m} h_{k,m,n},
\end{equation}
where $\beta_{k,m}\in\{0,1\}$ is the antenna activation indicator. Specifically, $\beta_{k,m}=1$ indicates that pinching antenna $m$ on waveguide $k$ is activated, while $\beta_{k,m}=0$ otherwise.
%%%%%%%%%%%%%%%%%%%%%%%%%%%%%%%%%%%%%%%%%%%%%%%%%
\subsection{Signal Model}
At the base station, the available transmit power $P_t$ is equally allocated among the $K$ waveguides. Accordingly, the transmitted signal fed into waveguide $k$ is expressed as
\begin{equation}
x_k=\sum_{n=1}^N\sqrt{\alpha_{k,n}\frac{P_t}{K}}s_n,
\end{equation}
where $\alpha_{k,n}\in\{0,1\}$ is the waveguide assignment indicator, and $s_n$ is the desired signal of user $n$. Specifically, $\alpha_{k,n}=1$ indicates that user $n$ is assigned to waveguide $k$, and $\alpha_{k,n}=0$ otherwise. It is noted that power loss within the dielectric waveguides is neglected, as its impact is negligible \cite{kaidi2025pin}.

At user $n$, the signals transmitted through all waveguides can be received and are presented as follows:
\begin{equation}
y_n=\sum_{k=1}^Kh_{k,n}\sqrt{\frac{P_t}{K}}\sum_{i=1}^N\sqrt{\alpha_{k,i}}s_i+z_n,
\end{equation}
where $z_n$ is the additive white Gaussian noise. The achievable data rate of user $n$ is then given by
\begin{equation}
R_n=\log_2\!\left(1\!+\!\frac{\sum_{k=1}^K\alpha_{k,n}\frac{P_t}{K}|h_{k,n}|^2}{\sum_{k=1}^K\sum_{i=1,i\neq n}^N\alpha_{k,i}\frac{P_t}{K}|h_{k,n}|^2+\sigma^2}\!\right),
\end{equation}
where $\sigma^2$ is the noise power.
%%%%%%%%%%%%%%%%%%%%%%%%%%%%%%%%%%%%%%%%%%%%%%%%%
\subsection{Problem Formulation}
To enhance the performance of the considered pinching-antenna system under LoS blockages, a sum rate maximization problem can be formulated by jointly optimizing waveguide assignment and antenna activation, as follows:
\begin{subequations}
\begin{empheq}{align}
\max_{\boldsymbol{\alpha},\boldsymbol{\beta}}\quad & \sum_{n=1}^N R_n\\
\textrm{s.t.} \quad & \alpha_{k,n}\in\{0,1\}, \forall k\in\mathcal{K}, \forall n\in\mathcal{N},\\
& \sum\nolimits_{k=1}^K\alpha_{k,n}=1, \forall n\in\mathcal{N},\\
& \sum\nolimits_{n=1}^N\alpha_{k,n}=1, \forall k\in\mathcal{K},\\
& \beta_{k,m}\in\{0,1\}, \forall m\in\mathcal{M}_k, \forall k\in\mathcal{K},\\
& \sum\nolimits_{m=1}^M\beta_{k,m}= 1, \forall k\in\mathcal{K},
\end{empheq}
\label{problem}
\end{subequations}\vspace{-2mm}\\
where $\boldsymbol{\alpha}$ and $\boldsymbol{\beta}$ are the collections of all waveguide assignment indicators and antenna activation indicators, respectively. Constraints (\ref{problem}c) and (\ref{problem}d) guarantee that each user is assigned to one waveguide, and each waveguide serves one user, respectively. Constraint (\ref{problem}f) ensures that one pinching antenna is activated on each waveguide.
%%%%%%%%%%%%%%%%%%%%%%%%%%%%%%%%%%%%%%%%%%%%%%%%%
%%%%%%%%%%%%%%%%%%%%%%%%%%%%%%%%%%%%%%%%%%%%%%%%%
\section{Matching based Waveguide Assignment and Antenna Activation}
Since the formulated problem is an integer programming problem, this section employs matching theory to derive a solution. By constructing different preference lists, a matching based algorithm is developed to provide a trade-off between performance and complexity.
%%%%%%%%%%%%%%%%%%%%%%%%%%%%%%%%%%%%%%%%%%%%%%%%%
\subsection{Construction of Three-Sided Matching}
In problem \eqr{problem}, users are assigned to waveguides, and pinching antennas are activated to serve the assigned users. This process can be modeled as a three-sided matching involving the sets of users, waveguides, and pinching antennas. In this matching, each user $n$ is matched with a waveguide $k$ and an antenna $m$, resulting in an one-to-one matching triple $(n,k,m)\in\mathcal{N}\times\mathcal{K}\times\mathcal{M}$, where $\mathcal{M}$ is the set of all pinching antennas. Moreover, there is a hierarchical relationship between antennas and waveguides, such that the activated antenna must belong to the corresponding waveguide, i.e., $m\in\mathcal{M}_k$. This matching is defined as follows:
\begin{definition}
A three-sided one-to-one matching $\Psi$ is a mapping function $\Psi:\mathcal{N}\rightarrow\mathcal{K}\rightarrow\mathcal{M}$ such that:
\begin{enumerate}[label=(\theenumi)]
\item $\Psi(n)=(k,m),\forall n\in\mathcal{N}$, $\Psi(k)=(n,m),\forall k\in\mathcal{K}$;
\item $\Psi(n)=(k,m)\Leftrightarrow\Psi(k)=(n,m)$.
\end{enumerate}
\end{definition}
The above matching formulation is derived from the constraints specified in problem \eqr{problem}, where the symbol $\Psi$ has different interpretations depending on the associated parameters.
%%%%%%%%%%%%%%%%%%%%%%%%%%%%%%%%%%%%%%%%%%%%%%%%%
\subsection{Design of Matching based Algorithm}
In the considered system, a user can only be reassigned to a different waveguide through a swap operation with another user, as defined by the concept of swap matching:
\begin{definition}
Given a matching $\Psi$, a swap matching between users $n$ and $n'$ is denoted by
\begin{equation}
\Psi_{n\leftrightarrow n'}\!=\!\Psi\backslash\{(n,k,m),(n',k',m')\}\cup\{(n,k',m'),(n',k,m)\},
\end{equation}
where $\Psi(n)=(k,m)$ and $\Psi(n')=(k',m')$.
\end{definition}

Based on the objective function of problem \eqref{problem}, the preference of any user is determined by the resulting sum rate. Specifically, if user $n$ prefers to be assigned to a different waveguide $k'$, the strict preference is expressed as follows:
\begin{equation}\label{pren}
(k,m, \Psi) \prec_n (k',m,\Psi_{n\leftrightarrow n'}) \Leftrightarrow \sum_{i=1}^N\!R_i(\Psi) < \sum_{i=1}^N\!R_i(\Psi_{n\leftrightarrow n'}),
\end{equation}
where $R_i(\Psi)$ is the data rate of user $i$ under matching $\Psi$. When users $n$ and $n'$ swap their assigned waveguides $k$ and $k'$, the channels for transmitting interference and desired signals are exchanged. As a result, the antenna activation strategy for the involved waveguides tends to be updated to adapt to the new waveguide assignment configuration. Specifically, if user $n$ is assigned to waveguide $k'$, the antenna activation strategy can be determined by considering the joint preference of the user and waveguide over all available pinching antennas, i.e.,
\begin{equation}\label{prek}
(m',\Psi_{n\leftrightarrow n'}) \prec_{(n,k')} (m'',\Psi_{n\leftrightarrow n'}')\Leftrightarrow \sum_{i=1}^N R_i(\Psi_{n\leftrightarrow n'}) < \sum_{i=1}^N R_i(\Psi_{n\leftrightarrow n'}').
\end{equation}
The above preference indicates that user $n$ and waveguide $k'$ tends to activate a different pinching antenna $m''$, instead of the original activated antenna $m'$. In this case, the matching is further transformed from $\Psi_{n\leftrightarrow n'}$ to $\Psi_{n\leftrightarrow n'}'$ as follows:
\begin{equation}
\Psi_{n\leftrightarrow n'}'=\Psi_{n\leftrightarrow n'}\backslash(n,k',m')\cup(n,k',m'').
\end{equation}

It can be observed from \eqr{pren} and \eqr{prek} that the transformation of the matching is driven by users. That is, users swap their assigned waveguides, which subsequently leads to updates in antenna activation. Therefore, a stable matching can be achieved by identifying swap-blocking pairs of users \cite{danilov2003existence}. The concept of the core is introduced to characterize stable matchings, defined as follows:
\begin{definition}\label{core}
A matching $\Psi$ is in the core if and only if there exists no swap-blocking pair $\langle n, n'\rangle$ such that user $n$ satisfies
\begin{equation}
(k,m,\Psi) \prec_n (k',m'',\Psi_{n\leftrightarrow n'}'), \forall m''\in\mathcal{M}_{k'},
\end{equation}
where $\Psi(n)=(k,m)$ and $\Psi(n')=(k',m')$.
\end{definition}
This definition implies that a swap-blocking pair exists if user $n$ has an incentive to swap the assigned waveguide from $k$ to $k'$ and change the activated pinching antenna from $m'$ to some $m''$. Since all involved players share the same utility function, the swap-blocking pair can be generated based on the the preference of either involved user.

\begin{algorithm}[t]
\caption{Matching based Algorithm}
\label{alg}
\begin{algorithmic}[1]
\FOR{$n\in\mathcal{N}$, where $\Psi(n)=(k,m)$}
\FOR{$k'\in\mathcal{K}$, where $\Psi(k')=(n',m')$}
\IF{$\langle n, n'\rangle$, where $n \neq n'$}
\STATE Obtain $\Psi_{n\leftrightarrow n'}$.
\FOR{$\kappa\in \{k ,k'\}$, where $\Psi_{n\leftrightarrow n'}(\kappa)=(\nu,\mu)$}
\FOR{$\mu'\in\mathcal{M}_{\kappa}$, where $\mu'\neq \mu$}
\STATE Obtain  $\Psi_{n\leftrightarrow n'}'$.
\IF{$(\mu,\Psi_{n\leftrightarrow n'}) \prec_{(\nu,\kappa)} (\mu',\Psi_{n\leftrightarrow n'}')$}
\STATE $\Psi_{n\leftrightarrow n'}=\Psi_{n\leftrightarrow n'}'$.
\ENDIF
\ENDFOR
\ENDFOR
\STATE $\Psi=\Psi_{n\leftrightarrow n'}$.
\ENDIF 
\ENDFOR
\ENDFOR
\end{algorithmic}
\end{algorithm}

Based on the preference of users and waveguides, a waveguide assignment and antenna activation algorithm is proposed in Algorithm \ref{alg}. Specifically, during initialization, an initial matching $\Psi$ is generated by randomly assigning users to waveguides and activating one pinching antenna on each waveguide. Whenever a swap-blocking pair is detected, the antenna activation strategy of the involved waveguides is updated accordingly (lines 5-12). This algorithm repeats until no swap-blocking pairs can be identified throughout a complete iteration over all users.
%%%%%%%%%%%%%%%%%%%%%%%%%%%%%%%%%%%%%%%%%%%%%%%%%
\subsection{Alternative Preference Design based on LoS and Distance}
In the considered pinching-antenna system with LoS blockages, the LoS path component dominates the sum rate, with the distance between users and antennas playing a critical role in determining the LoS path strength. Based on this characteristic, the preferences introduced in the previous subsection, which were originally based on the sum rate, can be reconstructed by exploiting the LoS indicators and distances.

For antenna activation, the activated pinching antenna on any given waveguide is switched only if the sum rate can be strictly increased, as shown in \eqr{prek}. This condition typically satisfied under the following scenarios:
\begin{subequations}
\begin{empheq}[left=\empheqlbrace]{align}
&\phi_{k',m''}^n> \phi_{k',m'}^n,\\
&\Delta D_{k',m'\to k',m''}^n> 0,\\
&\Phi_{k',m''}^n<\Phi_{k',m'}^n,
\end{empheq}
\label{prekre}
\end{subequations}\vspace{-2mm}\\
where $\Delta D_{k',m'\to k',m''}^n$ is the change in distance between user $n$ and the pinching antennas on waveguide $k'$, i.e.,
\begin{equation}
\Delta D_{k',m'\to k',m''}^n=\|\boldsymbol{\psi}_n-\boldsymbol{\psi}_{k',m'}^\mathrm{pin}\|-\|\boldsymbol{\psi}_n-\boldsymbol{\psi}_{k',m''}^\mathrm{pin}\|,
\end{equation}
and $\Phi_{k',m'}^n=\sum_{i=1,i\neq n}^N\phi_{k',m'}^i$ is the total number of LoS interference links from antenna $m'$ on waveguide $k'$ to all unassociated users. Inequality (\ref{prekre}a) implies that activating antenna $m''$ establishes a new LoS link between user $n$ and waveguide $k'$. Inequality (\ref{prekre}b) ensures that user $n$ is closer to the new antenna $m''$ than to the previous one $m'$. Inequality (\ref{prekre}c) requires that switching to antenna $m''$ reduces the number of LoS interference links. The preference in \eqr{prek} holds if at least one of the three conditions in \eqref{prekre} is strictly satisfied, while the other two hold with non-strict inequalities.

In terms of waveguide assignment, the preference involves two users and may potentially lead to antenna activation. As a result, the two users involved in the swap operation in \eqref{pren} are considered jointly, and the following conditions are derived:
\begin{subequations}
\begin{empheq}[left=\empheqlbrace]{align}
&\phi_{k',m'}^n+\phi_{k,m}^{n'}>\phi_{k,m}^n+\phi_{k',m'}^{n'},\\
&\Delta D_{k,m\to k',m'}^n+\Delta D_{k',m'\to k,m}^{n'}>0,\\
&\Phi_{k',m'}^n+\Phi_{k,m}^{n'}<\Phi_{k,m}^n+\Phi_{k',m'}^{n'}.
\end{empheq}
\label{preure}
\end{subequations}\vspace{-2mm}\\
It is worth noting that swapping waveguides does not affect uninvolved users, since their interference channels remain unchanged. However, waveguide assignment may trigger antenna activation. Therefore, inequality (\ref{preure}c) is included to ensure that the number of LoS interference links is strictly reduced. Due to this potential reactivation, the antennas $m$ and $m'$ in the above expressions may represent the newly activated pinching antennas resulting from the swap, rather than the originally activated ones. Similarly, when identifying a swap-blocking pair, at least one of these inequalities should be strictly satisfied, while the other two are satisfied non-strictly.

%%%%%%%%%%%%%%%%%%%%%%%%%%%%%%%%%%%%%%%%%%%%%%%%%
\subsection{Complexity, Convergence, and Stability}
The computational complexity of Algorithm \ref{alg} is approximately $\mathcal{O}(2CNKM)$, where $C$ is the total number of required cycles. This complexity arises because each round involves $N(K-1)$ swap operations, and each swap operation involves $2M$ antenna activation operations, as described in lines 5–12. It is worth pointing out that although the overall algorithm complexity remains the same, exploiting the LoS indicator and distance based preference can significantly reduce the actual computational load by avoiding the calculations of effective channels ($NK$ computations) and data rates ($N$ computations). Consequently, the total number of computations for Algorithm \ref{alg} with the LoS indicator and distance based preference is reduced to $\frac{1}{N(K+1)}$ of that required by the sum rate based preference.

The preference is defined based on strict criteria, including an increase in the sum rate, an increase in LoS links for desired signals, a decrease in LoS links for interference signals, and a reduction in the distance between the user and the activated antenna on the associated waveguide. Under these conditions, and given that the numbers of users, waveguides, and antennas are finite, the proposed matching based algorithm is guaranteed to converge. Moreover, since Algorithm \ref{alg} terminates only when no swap-blocking pairs can be identified in a complete cycle, the resulting matching is, by Definition~\ref{core}, always stable. In this case, no user or waveguide has an incentive to deviate from its current configuration.
%%%%%%%%%%%%%%%%%%%%%%%%%%%%%%%%%%%%%%%%%%%%%%%%%
%%%%%%%%%%%%%%%%%%%%%%%%%%%%%%%%%%%%%%%%%%%%%%%%%
\section{Simulation Results}
\begin{figure}[!t]
\hspace{-4mm}\centering{\includegraphics[width=100mm]{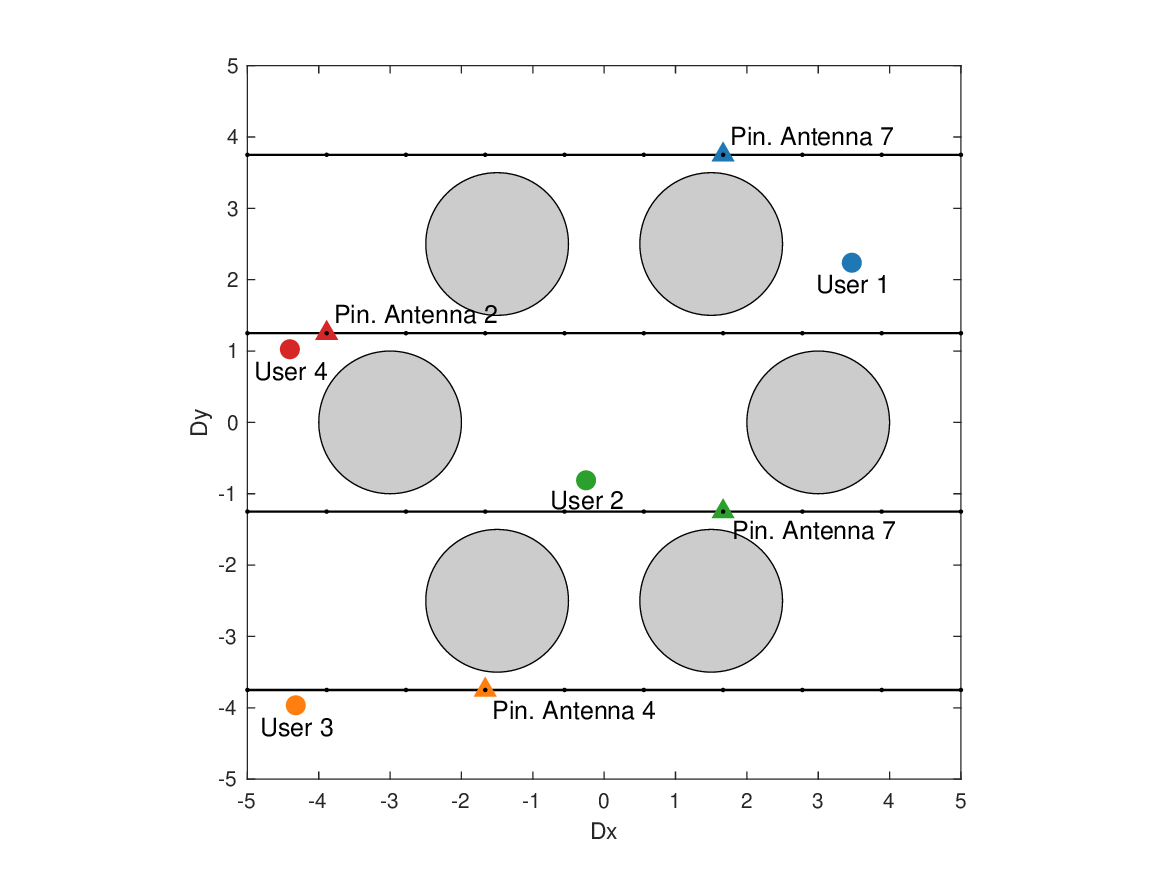}}
\caption{An illustration of the proposed solution, where $M=10$, $L=3$, $P_t=20$~dBm, and $R_B=1$~meters.}
\label{result1}
\vspace{-6mm}
\end{figure}

In this section, simulation results are presented to evaluate the performance of the proposed system and algorithm. The parameters are configured as follows: $K=N=4$, $D_x=D_y=10$~meters, $d=3$~meters, $f_c=28$~GHz, $\delta_\ell\sim\mathcal{CN}(0,1)$, and $\sigma^2=-80$~dBm. The waveguides are deployed in parallel at positions $y_n^\mathrm{Pin}=\pm 0.125$ and $y_n^\mathrm{Pin}=\pm 0.375$. To model the obstructed environment, $6$ blockages, each with a height of $3$ meters and a radius of $R_B$, are placed at fixed positions, as illustrated in \fref{result1}, while scatterers are randomly distributed within the rectangular region. The sum rate based preference and LoS and distance based preference correspond to Solution 1 and Solution 2, respectively. The conventional system with fixed-location antennas is included as a benchmark, in which $K$ antennas are placed at the center.

A representative case of waveguide assignment and antenna activation is illustrated in \fref{result1}, where the users assigned to the waveguides from top to bottom are $1$, $4$, $2$, and $3$, respectively. It shows that the proposed algorithm improves both system throughput and fairness in the pinching-antenna system with LoS blockages. In particular, the resulting strategy dynamically adapts to user locations to establish LoS links for desired signals while preserving NLoS links for interference signals, thereby simultaneously achieving both signal enhancement and interference suppression.

\begin{figure}[!t]
\centering{
\subfigure[$L=0$]{\centering{\includegraphics[width=80mm]{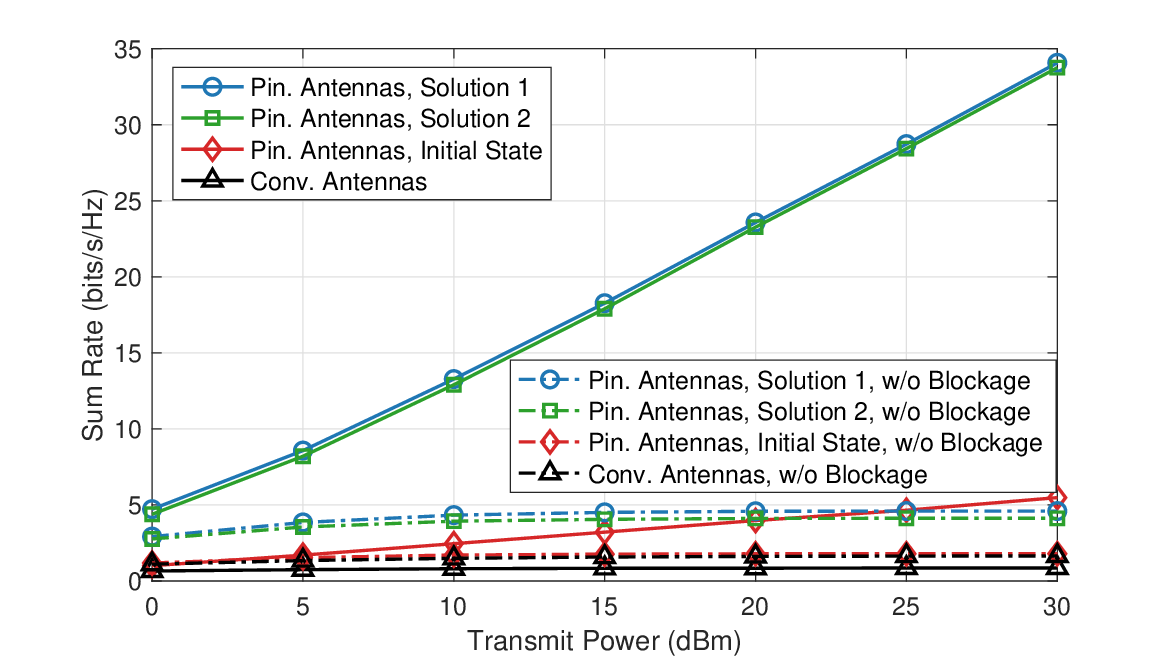}}}\vspace{-2mm}
\subfigure[$L=3$]{\centering{\includegraphics[width=80mm]{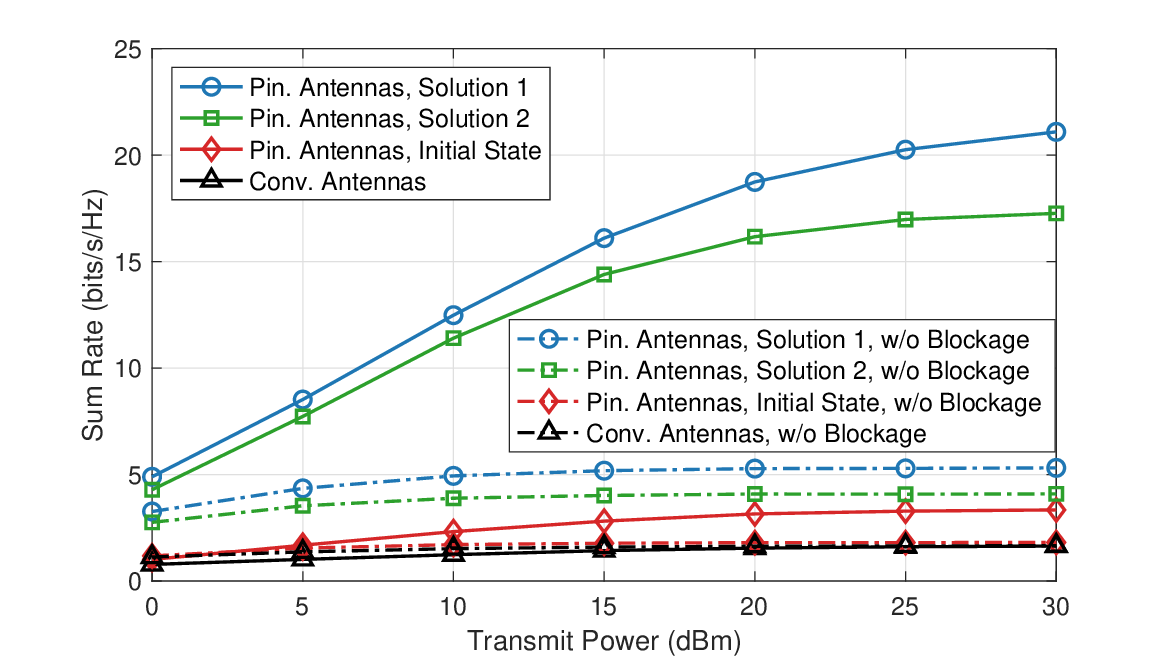}}}}\vspace{-2mm}
\caption{Impact of the transmit power on the sum rate, where $M=20$, and $R_B=1$~meters.}
\label{result2}
\vspace{-6mm}
\end{figure}

In \fref{result2}, the impact of transmit power is evaluated for scenarios with and without NLoS links. The results show that the proposed matching based algorithm can effectively exploit LoS blockages to suppress interference, thereby significantly enhancing the performance of the pinching-antenna system. Notably, even without optimization, pinching-antenna systems can outperform fixed-location antennas, particularly in obstructed environments. In the absence of blockages and at high transmit power, the signal-to-interference-plus-noise ratio (SINR) approaches a constant (e.g., $1/(N-1)$ in the initial state), resulting in a flat performance curve. Moreover, the LoS and distance based preference can achieve performance close to that of the sum rate based preference when scatterers are absent. However, since Solution 2 ignores the quality of NLoS links, the achievable sum rate achieved is lower than that of Solution 1 when scatterers are present. Nevertheless, due to its significant reduction in complexity, Solution 2 provides a practical trade-off between performance and efficiency.

\begin{figure}[!t]
\centering{\includegraphics[width=85mm]{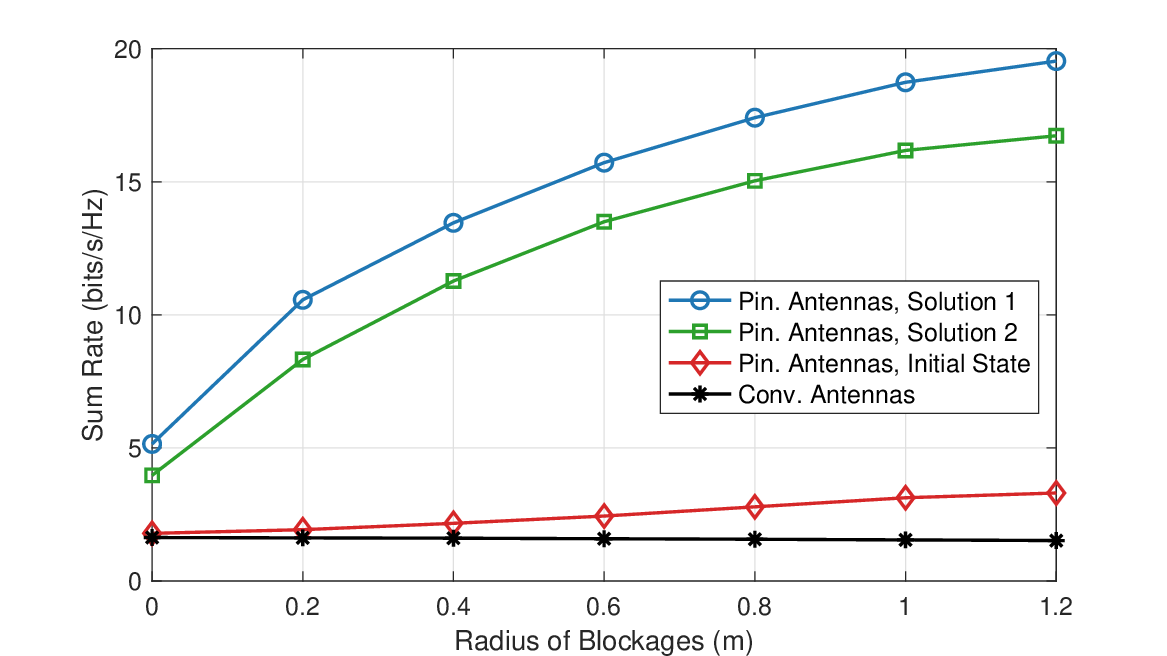}}
\caption{Impact of the radius of blockage on the sum rate, where $M=20$, $L=3$, and $P_t=20$~dBm.}
\label{result3}
\vspace{-6mm}
\end{figure}

\begin{figure}[!t]
\centering{\includegraphics[width=85mm]{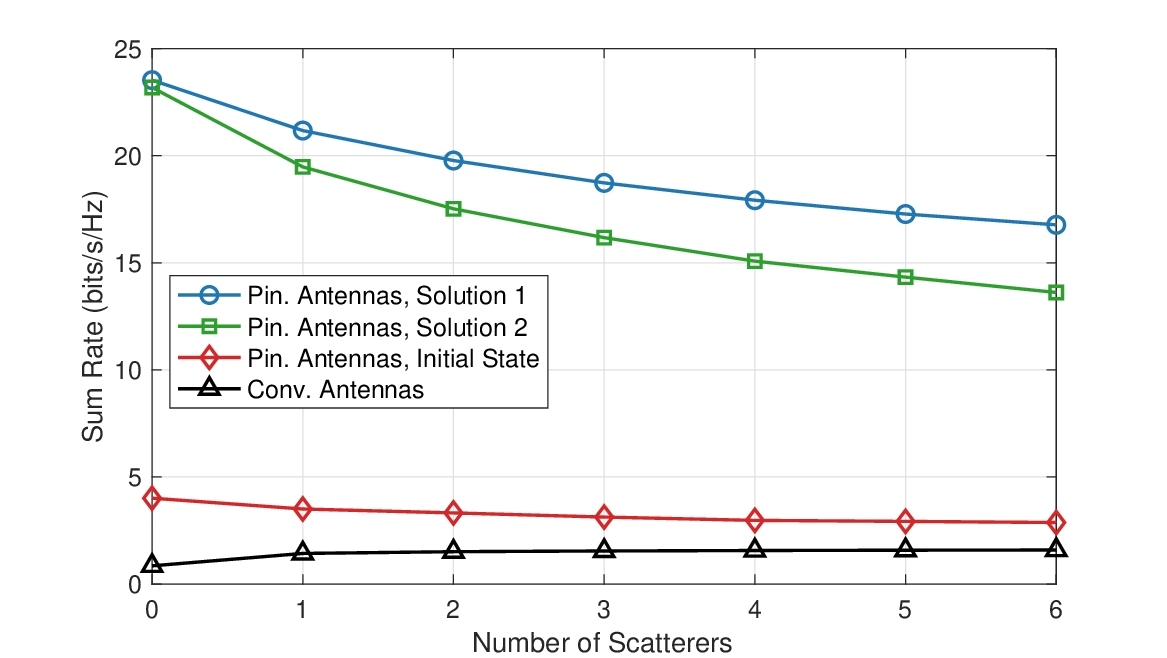}}
\caption{Impact of the number of scatterers on the sum rate, where $M=20$, $P_t=20$~dBm, and $R_B=1$~meters.}
\label{result4}
\vspace{-6mm}
\end{figure}

\fref{result3} illustrates the impact of blockage radius. As the blockage radius increases, the ability of the proposed solution to exploit LoS blockages for interference suppression improves, resulting in an enhanced achievable sum rate for the pinching-antenna system. In contrast, in the conventional fixed-location antenna system, a larger blockage radius degrades the effective channels, causing a monotonic decrease in the sum rate. The impact of the number of scatterers is presented in \fref{result4}. This parameter affects pinching-antenna and fixed-location antenna systems differently. In pinching-antenna systems, an increasing number of scatterers introduces more severe interference, which in turn reduces the achievable sum rate. For fixed-location antenna systems, however, additional scatterers enhance both desired and interfering signals simultaneously. As a result, the system performance is more sensitive to the presence or absence of scatterers rather than their specific quantity.
%%%%%%%%%%%%%%%%%%%%%%%%%%%%%%%%%%%%%%%%%%%%%%%%%
%%%%%%%%%%%%%%%%%%%%%%%%%%%%%%%%%%%%%%%%%%%%%%%%%
\section{Conclusions}
This work investigated pinching-antenna systems in the presence of LoS blockages, for which a mixed LoS and NLoS channel model was developed. Based on waveguide assignment and antenna activation, a sum rate maximization problem was formulated, and a matching based algorithm was proposed incorporating two different preference designs. Simulation results demonstrated that the pinching-antenna system achieves significant performance gains in obstructed environments, and the proposed algorithm effectively exploits LoS blockages for interference suppression.
%%%%%%%%%%%%%%%%%%%%%%%%%%%%%%%%%%%%%%%%%%%%%%%%%
%%%%%%%%%%%%%%%%%%%%%%%%%%%%%%%%%%%%%%%%%%%%%%%%%
\bibliographystyle{IEEEtran}
\bibliography{KaidisBib}
%%%%%%%%%%%%%%%%%%%%%%%%%%%%%%%%%%%%%%%%%%%%%%%%%
%%%%%%%%%%%%%%%%%%%%%%%%%%%%%%%%%%%%%%%%%%%%%%%%%
\end{document}